\documentclass[12pt]{iopart}
\usepackage{graphicx}
\usepackage{bm}
\usepackage{amssymb}
\usepackage{amsthm}
\newcommand{\be}{\begin{equation}}
\newcommand{\ee}{\end{equation}}
\newcommand{\bea}{\begin{eqnarray}}

\newcommand{\eea}{\end{eqnarray}}

\def\cR{{\cal R}}

\begin{document}
\title{Constraining $f(R)$ gravity in the Palatini formalism}
\author{Thomas P.~Sotiriou\footnote[1]{sotiriou@sissa.it}}

\address{SISSA-International School of Advanced Studies, via Beirut 2-4, 34014, Trieste, Italy}
\address{INFN, Sezione di Trieste, Via Valerio 2, 34014, Trieste, Italy}

\begin{abstract}
Although several models of $f(R)$ theories of gravity within the Palatini approach have been studied already, the interest was concentrated on those that have an effect on the late-time evolution of the universe, by the inclusion for example of terms inversely proportional to the scalar curvature in the gravitational action. However, additional positive powers of the curvature also provide interesting early-time phenomenology, like inflation, and the presence of such terms in the action is equally, if not more, probable. In the present paper models with both additional positive and negative powers of the scalar curvature are studied. Their effect on the evolution of the universe is investigated for all cosmological eras, and various constraints are put on the extra terms in the actions. Additionally, we examine the extent to which the new terms in positive powers affect the late-time evolution of the universe and the related observables, which also determines our ability to probe their presence in the gravitational action.
\end{abstract}

\maketitle
\section{Introduction}
\label{sec:1}
General Relativity has been an extremely successful theory of gravity for almost 100 years now. It has passed all tests on 
scales relevant for the solar system. However, all of these tests are actually relevant to the Post-Newtonian 
regime and not to the full version of the theory. At the same time, during the last decades, several indications have appeared, leading to 
the thought that maybe General Relativity is the effective part of a more general theory
of gravitation, applicable only on solar system scales. Some of these indications are of theoretical origin and as an example one could name the difficulties in finding the quantum counterpart
of general relativity and the fact that it has not yet been possible to include it in any grand unification scheme. The others are of observational origin, like the fact that relativity cannot explain the flat rotation 
curves of galaxies without introducing unseen dark matter and, of course, the fact that it cannot explain the current accelerated expansion of the universe 
without the introduction of both dark matter and
dark energy.
Also, in order to explain these cosmological observations, one has not only to introduce these exotic componets, but also to assume that they sum up to 96\% percent of the total
energy content of the universe. This, of course, does not sound appealing at all, since the nature of these components still remains a mystery, and this
 has lead to 
several attempts to modify or extend General Relativity in different ways, addressing the dark matter \cite{mond,mond1,mond2,mond3,mond4,mond5} or the 
dark energy \cite{faraoni,dvali,def,dvali2,ark} problem.

An alternative route is to modify General Relativity by abandoning the simplicity assumption that the action should be
linear in the scalar curvature  $R$. So, instead of using the Einstein--Hilbert Lagrangian one can use a more general one, which
depends on a generic function $f(R)$ (see \cite{buh,bar1,bar2} for early studies). Hence, such theories are called $f(R)$ theories of gravity. Such modifications are not completely arbitrary as higher
order terms in curvature invariants seem to be present in the effective Lagrangian of the gravitational field when
quantum corrections or string/M-theory are considered \cite{quant1,quant2,quant3,quant4,noji2,vassi}. 
Therefore, $f(R)$ theories seem also to address the theoretical concerns about General Relativity that were mentioned earlier.
Phenomenological interest up until recently was restricted to the inflationary behaviour exhibited when positive
powers of $R$ are present in the Lagrangian \cite{staro}. However, it was seriously increased when it was shown that a 
term inversely proportional to $R$ can lead to late time expansion \cite{capo, capo2,carroll}. Such theories, appealing as they may be, are unfortunately not free
of problems. First of all they lead to fourth order differential equations which are difficult to attack. Additionally, it is doubtful whether they
can pass the known solar system tests \cite{chiba,cem} or whether they have the correct 
Newtonian limit \cite{cem,dick,olmo,sot1}. It is possible that
more sophisticated models may exhibit behaviour closer to the expected one \cite{noji}, but this requires significant fine tuning of the 
various parameters. The most important problem of these models, however, is that they lead to unavoidable instabilities within matter
in a weak gravity regime \cite{dolgov}.

There is a further modification of gravity that one can consider which does not necessarily involve any modification
of the action, but rather the use of a different variational principle. This variational principle, known as the Palatini
formalism due to a historical misconception, since it was not introduced by Palatini but by Einstein himself, treats the metric and the affine
connection as independent geometrical quantities. One has to vary the Lagrangian with respect to both of them to derive the field equations, in contrast with the standard metric variation, where the Lagrangian is varied with respect to the metric alone, and the connections are assumed to be the Christoffel symbols of this metric. It is also known that 
when the Einstein-Hilbert action is used, the Palatini variational principle leads to the Einstein equations, just like the standard metric variation \cite{wald}. This is not true, however,
for a more general action. 

When used together with an $f(R)$ Lagrangian, the Palatini formalism leads to second order differential equations instead of the
fourth order ones that one gets with the metric variation. At the same time,
in vacuum, they straightforwardly reduce to standard General Relativity plus a cosmological constant \cite{ferr}. This ensures us that, firstly, the theory passes the solar system tests, and secondly,
that interesting aspects of GR like static black holes and gravitational waves are still present.
It is also free of the instabilities discovered in \cite{dolgov} for $f(R)$ gravity in the metric approach, which arise within matter in a weak gravity regime. Even though there was initially
some debate concerning the Newtonian limit \cite{meng2,barraco}, a recent paper \cite{sot1} seems to have settled this matter, showing that these theories indeed have the correct behaviour.
One more reason to study such a metric-affine variation is theoretical completeness. Both
approaches give exactly the same result when applied to the Einstein--Hilbert action. Thus there is no real criterion so far about which one of them
is better to use. Additionally, though, the Palatini variation seems to be more general since it yields GR without the need to pre-specify the 
relation between the metric and the connections. Finally it should be mentioned that the Palatini formalism is also closer
to the picture of a Hilbert space, since is assumes that there are two sets of independent variables, and therefore seems
more appealing when trying to quantize gravity.

In \cite{vollick}, Vollick showed that using the action of \cite{carroll}, which includes an $R^{-1}$ term, together with the Palatini variational
principle can lead to a theory of gravity that predicts late time accelerated expansion of the universe. Several models with similar behaviour followed (for a short review see \cite{meng3} and references therein).
Later it was proved that including an $R^2$ term in the action could not drive inflation \cite{meng} within this formalism. However, in \cite{sot2}
it was shown that this is not true for higher order terms in general, but merely that $R^2$  constitutes an exception
due to the construction of the theory, and a model was presented in which both an $R^3$ term and an $R^{-1}$ term were present in the action.
It was demonstrated that such a model can account both for early time inflation and late time
accelerated expansion.

It is important, of course, to go beyond the qualitative results and use the numerous
observations \cite{Riess,sloan,2dFGRS,wmap} to get quantitative ones.  Such a study was performed in \cite{ama}. Assuming that the gravitational
action includes, besides the standard linear term, a term inversely proportional to $R$, the authors used four different sets of cosmological data to
constrain it. These are the Supernovae type Ia gold set \cite{Riess}, the CMBR shift 
parameter \cite{bond}, the baryon oscillation length scale 
\cite{baryonsdss} and the linear growth factor at the 2dF Galaxy Redshift Survey effective redshift \cite{hawkins,tegmark}. However, as stated in the conclusions of \cite{ama}, the restricted form of $f(R)$, including only a term inversely proportional to $R$, prevents the study from being exhaustive.

Even though several models of $f(R)$ gravity in the Palatini formalism have been studied, most interest was concentrated on those having terms inversely proportional to the scalar curvature.
Bearing in mind that including an additional term proportional to a positive power of
the curvature in the action leads to interesting phenomenology \cite{sot2} and of course is equally, if not more, reasonable to have present in the action, here the behaviour of models with both positive and negative powers of $R$ will be investigated (we will be using the term positive powers of $R$ to imply positive powers additional to $R$ itself). By generalizing the results of \cite{sot2} it will be shown that such models lead to curvature driven inflation and late time cosmic expansion. However, a more detailed study reveals that such a curvature driven inflation is not interesting but is actually problematic, because there is no way to put an end to it, and should therefore be avoided. Another important question is how to constrain terms in positive powers of $R$. This problem will be addressed here in detail. One of the possible constraints will arise if we want to avoid the undesirable curvature driven inflation mentioned before, but also more standard ones, related to the later evolution and the Newtonian limit will be considered. The results will also be used to check whether the presence of such terms could affect the results of \cite{ama}. Additionally, a thorough discussion will take place, which can be considered relevant, not only to the results of this paper, but also to studies similar to \cite{ama}.

The rest of the paper is organized as follows: In section \ref{sec:2} the Palatini formalism is briefly reviewed. In section \ref{sec:3} a model including both negative and positive powers of the scalar curvature in the gravitational action is examined. Its behaviour is investigated for different cosmological eras and the possible constraints for the positive power term coming from different origins are obtained. Section \ref{sec:4} contains a thorough discussion of the physical consequences of the derived results. Also, the results of \cite{ama} are interpreted and discussed and we examine, using the results of section \ref{sec:3}, whether they can actually be consider general enough to include gravitational Lagrangians with positive powers in the scalar curvature as well.  Section \ref{sec:5} contains conclusions.

\section{ The Palatini formalism}
\label{sec:2}

We start by very briefly reviewing the Palatini formalism for a generalized action of the form
\be
S=\frac{1}{2\kappa}\int d^4 x\sqrt{-g}f(R)+S_M,
\ee
where $\kappa=8\pi G$ and $S_M$ is the matter action. For a detailed study see \cite{vollick}.

As mentioned in the introduction, within the Palatini formalism, one treats the metric and the affine connections as independent quantities. Therefore, we have to vary the action with respect to both of them in order to get the field equations. Varying with respect to the metric $g_{\mu\nu}$ gives
\be
\label{struct}
f'(R) R_{\mu\nu}-\frac{1}{2}f(R)g_{\mu\nu}=\kappa T_{\mu\nu},
\ee
where the prime denotes differentiation with respect to $R$ and the stress-energy tensor $T_{\mu\nu}$
is given by
\be
T_{\mu\nu}=-\frac{2}{\sqrt{-g}}\frac{\delta S_M}{\delta g^{\mu\nu}}.
\ee
Varying the action with respect to $\Gamma^{\lambda}_{\mu\nu}$ and contracting we get
\be
\label{ei1}
\nabla_{\alpha}[f'(R)\sqrt{-g}g^{\mu\nu}]=0,
\ee
where $\nabla$ denotes the covariant derivative with respect to the connections. It is straightforward now that the connections are the Christoffel symbols of the 
conformal metric $h_{\mu\nu}\equiv f'(R) g_{\mu\nu}$ (see \cite{vollick}). Thus,
we get
\be
\Gamma^{\lambda}_{\mu\nu}= \left\{^{\lambda}_{\mu\nu}\right\} +\frac{1}{2 f'}
\left[2\delta^{\lambda}_{(\mu}\partial_{\nu)}f'-g_{\mu\nu}g^{\lambda\sigma}\partial_{\sigma}f'\right],
\ee
where $\left\{^{\lambda}_{\mu\nu}\right\}$ denotes the Christoffel symbol of $g_{\mu\nu}$.
Finally, by contracting eq.~(\ref{struct}) one gets
\be
\label{scalar}
f'(R) R-2 f(R)=\kappa T.
\ee
Eq.~(\ref{scalar}) can in general be solved algebraically for $R=R(T)$, if of course the functional form of $f$ allows it.

If we are interested in cosmological solutions we can consider the spatially flat FLRW metric,
\be
\label{FRW}
ds^2=-dt^2+a(t)^2\delta_{ij}dx^idx^j.
\ee
and the perfect fluid energy-momentum tensor $T_\mu^{\nu}=\textrm{diag}(-\rho,p,p,p)$. The generalized Friedmann equation is given by
\be
\label{friedmann}
\left(H+\frac{1}{2} \frac{\dot{f'}}{f'}\right)^2=\frac {1}{6} \frac{\kappa (\rho+3p)}{f'}+\frac {1}{6}
\frac{f}{f'}.
\ee
We can use eq.~(\ref{scalar}) together with the conservation of energy to express $\dot{R}$ as a function of $R$ (see also \cite{sot2}):
\be
\label{rdot}
\dot{R}=-\frac{3H (R f'-2f)}{R f''-f'}.
\ee
Using eq.~(\ref{rdot}) to re-express $\dot{f'}(=f''\dot{R})$ and assuming that the universe is filled with dust ($p=0$) and radiation ($p=\rho/3$), after some mathematical manipulation, eq.~(\ref{friedmann}) gives
\be
\label{HRform}
H^2=
\frac{1}{6 f'}
\frac{2 \kappa \rho+R f'-f}
{\left(1-\frac{3}{2}\frac{f''(Rf'-2f)}{f'(Rf''-f')}\right)^2},
\ee
and $\rho=\rho_m+\rho_r$, where $\rho_m$ denotes the energy density of dust and $\rho_r$ the energy density of radiation.

\section{Cosmological behaviour and possible constraints}
\label{sec:3}

What we would like to investigate here is the cosmological behaviour of a model with both positive and negative powers of the scalar curvature in the gravitational action.
Such a general study seems to be a tedious analytical task given the complexity of the functions involved and the non-linearity of the equations. An alternative could be to do a numerical analysis which would however lead to lack of generality, and therefore should not be so appealing. Fortunately, there  seems to be a more elegant way to approach this problem, following the lines of \cite{sot2}. In some part of this paper we will leave the function $f$ unspecified and try to derive results independent of its form. However, in some cases we will also adopt the following representation for $f$, suitable for our purposes:
\be
\label{ans3}
f(R)=\frac{1}{\epsilon_1^{d-1}} R^d+R-\frac{\epsilon_2^{b+1}}{R^b},
\ee
with $\epsilon_1,\epsilon_2>0$, $d>1$ and $b\geq 0$; $b=0$ corresponds to the $\Lambda$CDM model when $\epsilon_1\rightarrow \infty$. The dimensions of $\epsilon_1$ and $\epsilon_2$ are $(\textrm{eV})^2$. Part of our task in this section will be to constrain the values of $\epsilon_1$. For the value of $\epsilon_2$ however, no extended discussion is really necessary, since we already know that in order for a model to be able to lead to late time accelerated expansion consistent with the current observations $\epsilon_2$ should by roughly of the order of $10^{-67}$ $(\textrm{eV})^2$.

In \cite{sot2} the specific case with $b=1$ and $d=3$ was discussed and it was shown that such a model can account for both early time inflation and late time accelerated expansion, and also give a smooth passage from one to the other during which standard cosmological eras can occur. Here we generalize this approach. First of all, let us see how a model with a general function $f$ would behave in vacuum, or in any other case that $T=0$ (radiation, etc.). 
If we define
\be
\label{defF}
F(R)\equiv f'(R)R-2f(R),
\ee
then eq.~(\ref{scalar}) reads
\be
\label{Fscalar}
F(R)=\kappa T,
\ee
and for $T=0$,
\be
\label{Fscalar0}
F(R)=0.
\ee
Eq.~(\ref{Fscalar0}) is an algebraic equation which, in general, will have a number of roots, $R_n$. Our notation implies that $R$ is positive so here we will consider the positive solutions (i.e. the positive roots). Each of these solutions corresponds to a de Sitter expansion, since $R$ is constant. If one wants to explain the late time accelerated expansion one of these solutions, say $R_2$, will have to be small. If in addition to this we also want our model to drive an early time inflation, there should be a second solution, $R_1$  corresponding to a larger value of $R$. 

For example, introducing in eq.~(\ref{scalar}) the ansatz given for $f$ in eq.~(\ref{ans3}), one gets
\be
\label{alg}
\frac{d-2}{\epsilon_1^{d-1}}R^{d+b}-R^{b+1}+(b+2)\epsilon_2^{b+1}=0.
\ee
If $\epsilon_1\gg \epsilon_2$ and $d>2$ then this equation has two obvious solutions
\be
R_1\sim \epsilon_1, \qquad R_2\sim \epsilon_2.
\ee
These solutions can act as seeds for a de Sitter expansion, since the expansion rate of the de Sitter universe scales like the square root of the scalar curvature. Notice that the rest of the solutions of eq.~(\ref{Fscalar0}), for $R>R_1$ and $R<R_2$ will not be relevant here. During the evolution we don't expect, as it will become even more obvious later on, that $R$ will exceed $R_2$ or become smaller than $R_1$.

\subsection{Early times}
\label{sec:31}
Following the lines of \cite{sot2} we can make the following observation. Since we expect the matter to be fully relativistic at very early times, $T=0$ and consequently $R$ is constant. This implies that the second term on the left hand side of eq.~(\ref{friedmann}) vanishes. Additionally, conservation of energy requires that the first term on the right hand side of the same equation scales like $a(t)^{-4}$, which means that the second term on the same side, depending only on the constant curvature, will soon dominate if $f(R)$ is large enough for this to happen before matter becomes non relativistic. $R$ can either be equal to $R_1$ or to $R_2$. Since we want $R_2$ to be the value that will provide the late time acceleration, $f$ should be chosen in such a way that $f(R_2)$ will become dominant only at late times when the energy densities of both matter and radiation have dropped significantly. Therefore, if we want to have an early inflationary era we have to choose the larger solution $R_1$, and $f$ should have a form that allows $f(R_1)$ to dominate with respect to radiation at very early times. The Hubble parameter will then be given by
\be
H\sim \sqrt{\frac{f(R_1)}{6 f'(R_1)}},
\ee
As an example we can use the ansatz given in eq.~(\ref{ans3}). The modified Friedmann equation reads
\be
H\sim \sqrt{\frac{\epsilon_1}{3 (d+1)}},
\ee
and the universe undergoes a de Sitter expansion which can account for the early time inflation.

\begin{figure}[h]
\begin{center}
\includegraphics[width=14cm,angle=0]{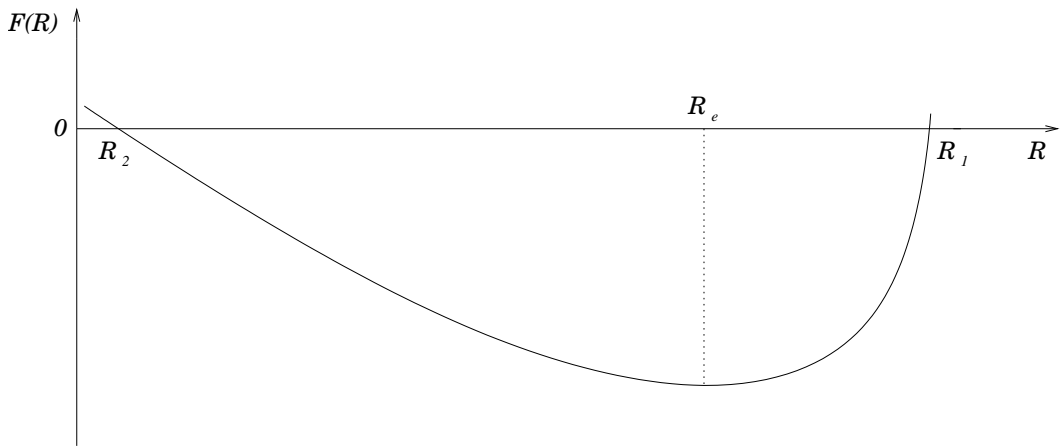}
\caption{The behaviour of a general function $F(R)$ over the interval $R_1>R>R_2$. $R_e$ denotes the value of $R$ where $F$ has a minimum. From this graph one can easily see that $F'$ is possitive when $R_e<R<R_1$ and negative when $R_2<R<R_e$.}\label{fig1}
\end{center}
\end{figure}

Sooner or later this inflationary expansion will lead to a decrease of the temperature and some portion of the matter will become non relativistic. This straightforwardly implies that $R$ will stop being constant and will have to evolve. $f'(R)$ plays the role of the conformal factor relating the two metrics $g_{\mu\nu}$ and $h_{\mu\nu}$, and therefore we do not consider sign change to be feasible throughout the evolution of the universe. We also know that in a certain range of values of $R$ it should be close to one. This is the case because there should be a range of values of $R$, for which $f(R)$ behaves essentially like $R$, i.e. our theory should drop to standard General Relativity, in order for us to be able to derive the correct Newtonian limit (see \cite{sot1} and section \ref{sec:33}). Together with $T\leq0$, the above implies the following:
\be
\label{constraints1}
f'(R)> 0, \quad F<0, \qquad \forall\quad R_2<R<R_1.
\ee
Since $F$  is a continuous function keeping the same sign in this interval and $F(R_1)=F(R_2)=0$, there should be a value for $R$, say $R_e$, where $F'(R_e)$=0, i.e. an extremum. Eq. (\ref{Fscalar}) implies that the time evolution of $R$ is given by
\be
\label{rdot1}
\dot{R}={\kappa \dot{T}}/{F'(R)}.
\ee
Differentiating eq.~(\ref{defF}) we get
\be
\label{above}
F'=f''R-f'.
\ee 
Using the fact that $\dot{f'}=f''\dot{R}$, and using eq.~(\ref{above}) to express $f''$ in terms of $F'$, $f'$ and $R$ one can easily show that
\be
\label{fdot}
\frac{\dot{f'}}{f'}=\frac{F'+f'}{R f'}\dot{R}.
\ee
The constraints given in eq.~(\ref{constraints1}) imply that for $R_e<R<R_1$, $F'>0$. An easy way to understand this is to remember that $F$ is negative in that interval but zero at $R=R_1$ and so it should be an increasing function (see also fig. \ref{fig1}). Since, $f'$ and $R$ are also positive, then what determines the sign of $\dot{f'}/{f'}$ in the neighbourhood of $R_e$ is the sign of $\dot{R}$. 

Let us see what will happen if we require $R$ to decrease, i.e. $\dot{R}<0$. Eq. (\ref{rdot}) implies that as $R\rightarrow R_e$, $\dot{R}\rightarrow -\infty$ if $\dot{T}\neq 0$, since $F'(R_e)=0$. Therefore, $\dot{f'}/{f'}\rightarrow -\infty$ and using eq.~(\ref{friedmann}) we can infer that $H\rightarrow \infty$. Physically, the above imply the following: $R$ has no way to decrease to a value less than $R_e$ without giving the universe an infinite expansion. In practice, any attempt for $R$ to approach $R_e$ would lead to a dramatically fast expansion, up until non-relativistic matter fully dilutes, and $R$ settles back to $R_1$. Thus, once the curvature terms in the modified Friedmann equation dominate the evolution, there is no turning back to matter domination through a continuous process. The two vacuum solutions $R_1$ and $R_2$ seem to be somehow disconnected in the evolution, and $R$ has to remain in the region close to only one of them. This is a general statement independent of the form of matter that is present, since $T$ was left unspecified in its derivation. So, even though, as shown in \cite{sot2}, including positive powers of $R$ in the action can lead to early time inflation, there seems to be no graceful exit from it. The only alternative left would be to consider that due to some other physical and non-classical process, the validity of the equation presented here ceases to hold for some time interval, which, however seems highly unphysical.

Since it seems impossible to provide an exit from this gravity driven inflation it seems reasonable to check if we can at least totally avoid it. If we choose as our initial solution $R_2$ instead of $R_1$ then the curvature terms will not dominate as long as $R$ is constant. However, there is still one subtle point. At some stage during the evolution the energy density of non-relativistic matter will have to rise sooner or later, forcing $R$ to change its value. It is also reasonable to assume that if inflation is not driven by curvature we will have to adopt a more standard approach to guarantee that it will happen, like an inflaton field. It is obvious, however, keeping in mind the previous discussion, that one would want $R$ to always be less than $R_e$ and this will impose a constraint which will depend on the functional form of $f$. For example if one assumes that $f$ is described by the ansatz given in eq.~(\ref{ans3}), then, considering ordinary matter, $R<R_e$ at all times implies that $\epsilon_1\gg\kappa\rho_m$ at all times. Let us also consider the case of a slow-rolling inflaton field, $\phi$. Then, if we denote its energy density by $\rho_\phi$, and its pressure by $p_\phi$, we have, as usual,
\bea
\rho_\phi&=&\frac{1}{2}\dot{\phi}^2+V(\phi),\\
p_\phi&=&\frac{1}{2}\dot{\phi}^2-V(\phi),
\eea
where $V(\phi)$ is the scalar field potential. During the period when $\phi$ dominates the evolution $T=\dot{\phi}^2-4 V(\phi)$, and since slow-roll implies that $\dot{\phi}^2\ll V(\phi)$, $T\approx -4 V(\phi)$. Therefore, if we want $R<R_e$, so that inflation proceeds as usual, then 
$\epsilon_1\gg V(\phi)$ at all times.

If $R$ is less than $R_e$ for all values of $T$ then it is easy to verify that everything will evolve naturally after the end of inflation.  For $R_2<R<R_e$, $F'<0$ and as non-relativistic matter dilutes, $\dot{\rho}_m<0$, so from eq.~(\ref{rdot1}) we see that $\dot{R}<0$ so that $R$ will have to decrease to reach the value $R_2$ asymptotically.

The above discussion, is not relevant of course if the only positive power present in the action is $R^2$. In this case, due to the form of $F$, this term does not appear in eq.~(\ref{Fscalar}) (see also \cite{sot2}). This specific case has been studied in \cite{mw}. One thing that is worth commenting upon, before closing this discussion, is the following. The constraint $f'>0$ (see eq.~(\ref{constraints1})), which is implied by the fact that $f'$ plays the role of the conformal factor relating the metrics $g_{\mu\nu}$ and $h_{\mu\nu}$, can, depending on the form of $f$ impose a further constraint on the value of the constants in front of the positive power terms. In \cite{mw2} inflation driven by an inflaton field was studied in the presence of an $R^2$ term. The authors derived a constraint for the constant appearing in front of the $R^2$ term in the action, by requiring that the square of the Hubble parameter is positive during the kinetic dominated phase. This constraint is exactly what would one derive by requiring $f'$ to be always positive.

\subsection{Big bang nucleosynthesis}
\label{sec:32}

Let us now turn our attention to the following cosmological era, radiation domination and big bang nucleosynthesis (BBN). Current observations indicate that the standard cosmological model can fit the data related to the primordial abundances of light elements. On the other hand, how a modified gravity model like the one discussed here would fit those data has not been yet worked out. However, there is little room for modifying the behaviour of the Friedmann equation during BBN and it seems reasonable to ask that the model under investigation should resemble standard cosmology during these standard cosmological eras \cite{bbn}. This implies that eq.~(\ref{HRform}) should be similar to the standard Friedmann equation
\be
\label{stfriedmann}
H^2=\frac{1}{3}\kappa \rho.
\ee
By comparing eqs. (\ref{HRform}) and (\ref{stfriedmann}) one see that during BBN
\be
\label{c1}
f'\sim 1
\ee
\be
\label{c2}
1-\frac{3}{2}\frac{f''(Rf'-2f)}{f'(Rf''-f')}\sim 1
\ee
\be
\label{c3}
R f'-3f\sim 0
\ee
To make the picture clearer we give the explicit expressions for $f'$ and $f''$ when $f$ is given by eq.~(\ref{ans3}):
\bea
f'&=&d \frac{R^{d-1}}{\epsilon_1^{d-1}}+1+b \frac{\epsilon_2^{b+1}}{R^{b+1}},\\
f''&=& d (d-1) \frac{R^{d-2}}{\epsilon_1^{d-1}}-b(b+1) \frac{\epsilon_2^{b+1}}{R^{b+2}}.
\eea
Let us for the moment assume that the term inversely proportional to $R$ is not present. In order for condition (\ref{c1}) to be fulfilled $\epsilon_1\gg R_{BBN}$. This is the natural constraint on the value of $\epsilon_1$ imposed when one asks for the model to have almost identical behaviour to the standard one during BBN. Once $\epsilon_1$ is chosen to have a large enough value all three constraints (\ref{c1}), (\ref{c2}) and (\ref{c3}) are easily fulfilled and the modified Friedmann equation (\ref{HRform}) becomes identical to the standard one, eq.~(\ref{stfriedmann}), for the relevant values of $R$. The above constraint can be viewed as a sufficient constraint for the model to be viable but not as a necessary one. However, one could also claim that, even if the modifications in the Friedmann equation do not necessarily have to be negligible, they should at least lead to second order corrections and not affect the leading order. This implies that $\epsilon_1$ should definitely be larger than $R_{BBN}$. 

We have, however, neglected the presence of the term inversely proportional to $R$. In the absence of positive powers this term should be negligible during BBN since $R_{BBN}$ is much larger than $\epsilon_2$. This picture may change if we consider the full version of the model. Eq.~(\ref{scalar}) can take the following form
\be
\label{red}
Rf'-2f=-\kappa \rho_m^0 (1+z)^3,
\ee
where $\rho_m^0$ is the present value of the energy density of non-relativistic matter and $z$ is the redshift. We have assumed here that $a_0=1$. Using eq. (\ref{red}) one can derive how $R$ will scale with the reshift. If $\epsilon_1\gg R_{BBN}$, then for all the evolution of the universe after BBN, $R$ scales almost like $(1+z)^3$. This indicates that, since BBN takes place at a very high redshift, $R_{BBN}$ is indeed much larger than $\epsilon_2$. If however, one assumes that $\epsilon_1$ is large enough to alter the behaviour of eq. (\ref{red}), then $R$ will have a milder scaling with the redshift, meaning that $R_{BBN}$ can get very close to $\epsilon_2$. Then the three constraints (\ref{c1}), (\ref{c2}) and (\ref{c3}) might not be fulfilled not only due to the presence of the positive power of $R$, but also because of the presence of the negative one. This will be a secondary effect related to the term with a positive power as shown earlier. It will be avoided if again $\epsilon_1\gg R_{BBN}$ and will be subdominant if $\epsilon_1$ is just smaller than $R_{BBN}$. Unfortunately, since the value of $R_{BBN}$ is very much model dependent it is difficult to turn this constraint into a numerical one.

\subsection{The Newtonian limit}
\label{sec:33}

Before, examining the behaviour of the models under discussion at late times we will first check their Newtonian limit. The formalism for doing this has been derived in \cite{sot1}. Since we are using a metric-affine variational principle, the connections are not the Christoffel symbols of the metric $g_{\mu\nu}$, but of the conformal metric
\be
h_{\mu\nu}=f'(R)g_{\mu\nu}.
\ee
Expressed in terms of $h_{\mu\nu}$ the field equations read
\be
\label{ein}
R_{\mu\nu}-\frac{1}{2}R h_{\mu\nu}+(f'-1)\left(R_{\mu\nu}-\frac{R}{2f'}h_{\mu\nu}\right)=\kappa T_{\mu\nu}.
\ee
It is easy to see that if $f'=1$, eq. (\ref{ein}) reduces to the Einstein equation and the metrics $h_{\mu\nu}$ and $g_{\mu\nu}$ coincide. Thus the model will have the correct Newtonian limit if $f'$ is very close to $1$ for curvatures of the order  of the solar system one. Current tests have an accuracy of $10^{-16}$ which implies that 
\be
R_{ss}< 10^{-16/(d-1)} \epsilon_1,
\ee
where $R_{ss}$ is the scalar curvature for the solar system density. We can use eq.(\ref{scalar}) to express the curvature in terms of the matter density. For the solar system $\rho\sim 10^{-11} \textrm{g/cm}^3$ and after some manipulations we get
\be
\label{Ncon}
\epsilon_1>10^{16/(d-1)-45} (\textrm{eV})^2.
\ee
In all the above we have neglected the contribution of the negative power of $R$ since its contribution to $f'$ for this range of densities is of the order of $10^{-21 (b+1)}$. See also \cite{sot1} for a more detailed discussion of the Newtonian limit. We also know that the current value of $R$, $R_0$, is of the order of $\epsilon_2$, so it easy to express eq. (\ref{Ncon}) in terms of $R_0$:
\be
\label{Ncon2}
\epsilon_1>10^{16/(d-1)+21} R_0.
\ee
This condition will prove useful later.

\subsection{Late times}
\label{sec:34}

Now let us check what the behaviour of the modified Friedmann equation will be at late times. The scalar curvature $R$ decreases with time to reach a value close to $\epsilon_2$. Therefore the conditions (\ref{c1}), (\ref{c2}) and (\ref{c3}) will at some point cease to hold because of the term involving the negative power of $R$. Any contribution of the term involving the positive power of $R$ will be negligible for two reasons. Firstly, since the value of $\epsilon_1$ should be such that these terms are already negligible during BBN, it is safe to assume that they will remain so all through the rest of the evolution of the universe. The same results can be inferred by using the constraints derived in section \ref{sec:31}. Secondly there is also the constraint on the value of $\epsilon_1$ coming from the Newtonian limit. As we mentioned earlier, the constraints coming from BBN are sufficient for a viable model but are not absolutely necessary, since it is not yet clear how well a model like that would fit the data from light element abundances. One could still assume that there is some slight contribution of the terms being discussed in the modified Friedmann equation during BBN, which in any case becomes even weaker at later times. The constraints coming from the early time behaviour are necessary but it is difficult to turn them into numerical ones. At the same time one can always claim that the early time evolution of the universe is not very well established and there might still be room for new physics there, affecting these constraints. However, the Newtonian constraint (\ref{Ncon}) is very straightforward and unquestionably necessary. This constraint will prove more than sufficient for our purposes in this section.

The range of values of $R$ which is of interest for late time observations is between the value of $R$ at decoupling $R_{dec}$ and $R_0$. Using eq.~(\ref{red}) it is easy to show that $R_{dec} (1+z_{dec})^{-3}\sim R_0$, or using as the value of the redshift at decoupling $z_{dec}=1088$ one gets $R_{dec}\sim 10^{10} R_0$. Eq.~(\ref{Ncon2}) then implies that $\epsilon_1>10^{16/(d-1)+11} R_{dec}$. Therefore the positive power term will be at least eleven orders of magnitude smaller than the linear term for any $d\geq 2$ and will become even more negligible as time passes. Thus, effectively
\bea
f&\sim& R-\frac{\epsilon_2^{b+1}}{R^b},\\
f'&\sim&1+b \frac{\epsilon_2^{b+1}}{R^{b+1}},\\
f''&\sim& -b(b+1) \frac{\epsilon_2^{b+1}}{R^{b+2}},
\eea
with extremely high accuracy for all times after decoupling. It is easy to see that the modified Friedmann equation of the model described in (\ref{ans3}) will be identical to that of a model with no positive powers of the curvature ($\epsilon_1\rightarrow \infty$) for late times.

In \cite{ama} the authors consider $f$ to be of the form
\be
\label{ans}
f(R)=R\left(1+\alpha\left(\frac{R}{H_0^2}\right)^{\beta-1}\right),
\ee
where $\alpha$ and $\beta$ are dimensionless parameters, with $\beta<1$ (note that in our notation
$R$ is positive). This representation of the function $f$ is very useful when one wants to constrain some dimensionless parameter. Comparing it with our ansatz, eq.~(\ref{ans3}), we get  $d=\delta+1$, $b=-\beta$ and $\epsilon_1 \rightarrow \infty$, since in eq.~(\ref{ans}) there is no positive power of $R$.
In order to constrain the values of  $\alpha$ and $\beta$ they use a rather extensive list of cosmological observations. The first quantity which they consider is the CMBR shift parameter \cite{bond,melchiorri,odman} which in a spatially 
flat universe is given by 
\be
\label{shiftdef}
\cR = \sqrt{\Omega_m H_0^2}\int_0^{z_{dec}}\frac{d\tilde{z}}{H(\tilde{z})},
\ee
where $z_{dec}$ is the redshift at decoupling and $\Omega_m\equiv \kappa \rho^0_m/(3H_0^2)$. When expressed in terms of the scalar curvature, eq.~(\ref{shiftdef}) becomes
\bea
\label{shift}
\cR & = & \sqrt{\Omega_m H_0^2}\int_{0}^{z_{dec}}\frac{dz}{H(z)}\nonumber\\
& = & \sqrt{\Omega_m H_0^2}\int_{R_{dec}}^{R_0}\frac{a'(R)}{a(R)^2}
\frac{dR}{H(R)}\\
& = & \frac{1}{3^{4/3}}\left(\Omega_m H_0^2\right)^{1/6}\int_{R_0}^{R_{dec}}
\frac{Rf''-f'}{\left(Rf'-2f\right)^{2/3}}\frac{dR}{H(R)}\nonumber.
\eea
Using the values for $z_{dec}$ and $\cR$ obtained with WMAP \cite{wmap}, namely 
$z_{dec}=1088^{+1}_{-2}$ and
$\cR=1.716\pm 0.062$, they find that the best fit model is$(\alpha,\beta)=(-8.4,-0.27)$.
They also use the ``Gold data set'' of Supernovae \cite{Riess}. What is important for this analysis is the expression of the luminosity distance, which in terms of $R$ is
\bea
\label{lumdist}
D_L(z) & = & (1+z)\int_{0}^z\frac{d\tilde{z}}{H(\tilde{z})}\nonumber\\
& = & \sqrt{\Omega_m H_0^2}\frac{1}{a(R)}\int_{R}^{R_0}\frac{a'(R)}{a(R)^2}\\
& = & \frac{1}{3}\sqrt{\Omega_m H_0^2}\left(Rf'-2f\right)^{1/3} \times \nonumber\\&&\int_{R_0}^{R_{dec}}
\frac{Rf''-f'}{\left(Rf'-2f\right)^{2/3}}\frac{dR}{H(R)}\nonumber.
\eea
Marginalizing over the Hubble parameter $h$, the authors again constrain $\alpha$ and $\beta$ and the best fit model is $(\alpha,\beta)=(-10.0,-0.51)$. Another independent observation which they use is that of the imprint of the primordial baryon-photon acoustic oscillations on the matter power spectrum. The dimensionless quantity $A$ \cite{linder,eisenstein,hu},
\begin{equation}
A= \sqrt{\Omega_m}E(z_1)^{-1/3}\left[ \frac{1}{z_1}\int_{0}^{z_1}\frac{dz}{E(z)}\right]^{2/3},
\end{equation}
where $E(z)=H(z)/H_0$
can act as
 a ``standard ruler''. The data from the Sloan Digital Sky Survey
\cite{baryonsdss} provide a value for $A$, namely
\be
\label{boe}
A=D_v(z=0.35)\frac{\sqrt{\Omega_m H_0^2}}{0.35c}=0.469\pm0.017,
\ee
where 
\be
\label{distance}
D_v(z)=\left[D_M(z)^2\frac{cz}{H(z)}\right]^{1/3},
\ee
and $D_M(z)$ is the comoving angular diameter distance. The best fit model using this value is
$(\alpha,\beta)=(-1.1,0.57)$. Finally, in \cite{ama} these three sets of data are combined to give a best fit for
$(\alpha,\beta)=(-3.6,0.09)$.

The above observations are potentially very useful, of course, in studying the viability of a model like (\ref{ans}). I will comment on the interesting results of \cite{ama} in the next section. However, as shown here, the modified Friedmann equation of a more general model like (\ref{ans3}), which also includes positive powers of $R$, is effectively identical to that of (\ref{ans}) at late times. Therefore, it is expected that the results of \cite{ama} will remain unaffected by the inclusion of positive powers of the scalar curvature, since these terms have to satisfy the constraints derived in this section.

Before closing this section I would like to briefly discuss the fourth scheme used in \cite{ama} in order to obtain constraints: large scale structure and growth of perturbations. The authors use 
the Jebsen-Birkoff theorem which, as they say, has not proved to hold for $f(R)$ theories in general. Of course one would probably expect that it does hold, and once one accepts their assumption, their results come naturally. However, keeping in mind that these results cannot actually improve the constraints obtained with the three schemes already mentioned, there is no real reason to pursue such a study further. Therefore in the present paper I will not comment further or try to generalize these results. As the authors of \cite{ama} correctly state, a more detailed analysis should be performed along the lines of \cite{tomi}.

\section{Discussion and physical interpretation of the constraints}
\label{sec:4}

It was shown that a model with a positive and a negative power of the scalar curvature added to the Einstein-Hilbert action can lead to the following evolutionary eras: a curvature driven inflation; a radiation dominated era in which BBN can take place, and during which the modified Friedmann equation can effectively be identical to the standard one, followed normally by a matter dominated era; a late-time era in which the universe undergoes a phase of accelerated expansion. This is in accordance with the results of \cite{sot2} and requires certain constraints. 

However, as shown in section \ref{sec:31}, a curvature driven inflation of this sort comes with serious problems. The curvature cannot drop to lower values in a continuous way without leading to an infinite value for the Hubble parameter. This implies that if the curvature is indeed dominant at early times, it is not possible to return to standard cosmology at later times. Therefore, if higher order terms are indeed present in the action, one has to assume that the initial value of $R$ is one that does not lead to early time curvature domination. Even in this case however, the theory comes together with a constraint. For a model described by eq.~(\ref{ans3}) this is $\epsilon_1\gg\rho_m$, which should hold through all of the evolution of the universe. If, additionally, one asks for an inflaton field inflation to occur, $\epsilon_1\gg V(\phi)$ at all times. These bounds imply that $\epsilon_1$ has to have a very large value, and are in accordance with the value of $\epsilon_1$ expected from the perspective of effective field theory, $\epsilon_1\sim M_p^2$. Even though the analysis that leads to these result does not apply for the exceptional case of $d=2$, as mentioned in section \ref{sec:31}, effectively the same constraints still have to hold, as shown in \cite{mw2}. As mentioned here, they can easily be inferred from the positivity of $f'$ during the kinetic domination phase of the inflaton. The above results indicate that the presence of positive powers of $R$ in the action cannot lead to interesting phenomenology and at the same time be in accordance with the established late time evolution of the universe. Such terms might be present in the action resulting from a more fundamental theory, but they would have to be very subdominant, even at early times and large curvatures.

For the Friedmann equation  not to be seriously modified during BBN we found that $\epsilon_1>R_{BBN}$, which is a model dependent constraint. Finally requiring our model to have the correct Newtonian limit gave $\epsilon_1>10^{16/(d-1)-45} (\textrm{eV})^2$ or in terms of the current value of the curvature on cosmological scales, $\epsilon_1>10^{16/(d-1)+21} R_0$. This last constraint is of course a necessary one, whereas the first one is related to the phenomenological behaviour that our model should have. The constraint coming from BBN is definitely sufficient for the model to be viable and seems to be a necessary one, but a more detailed study is needed to reach a final conclusion about this.

Making use of the derived constraints, and especially the one coming from the Newtonian limit, it has been shown that the presence of positive powers of the scalar curvature in the gravitational action cannot lead to any modification in the late time form of the Friedmann equation. Since 
this equation describes the evolution of the universe, the above statement implies that
the late evolution of the universe is {\it not} affected by the positive powers of the scalar curvature present in the action. This can be rephrased in two interesting ways: {\it The results of observational tests relevant to the late time evolution of the universe are insensitive to the inclusion of additional positive powers of $R$} or {\it observational tests relevant to the late time evolution cannot constrain the presence of additional positive powers of $R$ in the gravitational action}.
The first expression makes it clear that the conclusions drawn from \cite{ama}, or from any similar study, are actually more general than the authors claim. The second one however might be even more interesting since it implies that such tests are not sufficient to judge the overall form of the gravitational action. We will come to this shortly.

We now discuss the results of \cite{ama}, which were reviewed in section \ref{sec:2}. The best fit model for the combination of the different data sets suggests that their exponent $\beta$ is equal $0.09$ (see eq.~(\ref{ans})) and therefore favours the $\Lambda$CDM model, being well within the $1\sigma$ contour. However, one gets different values for $\beta$ when the different data sets are considered individually. For the SNe data the best fit model has $\beta=-0.51$ and the baryon oscillations $\beta=0.57$, both disfavouring the $\Lambda$CDM model, but also being mutually contradictory. The CMBR shift parameter gives $\beta=-0.27$ which again is significantly different from the other two values. Of course one might expect that the combination of the data will give the most trustworthy result. However, this is not necessarily true, and one could regard the discrepancies in the value of $\beta$ coming from different observation as an indication that more accurate data are needed to derive any safe conclusion. It would also be interesting to study to what extent a model with $\beta=-1$, which is the model commonly used in the literature, can individually fit the current data, and to compare the results with other models built to explain the current accelerated expansion, such as quintessence, scalar-tensor theories, etc. Bear in mind that the $\Lambda$CDM model has always been the best fit so far. However, the motivation for creating alternative models does not come from observations but from our inability to solve the theoretical problems that come up if one adopts the standard picture (coincidence problem, etc.).

Notice that the results derived in section \ref{sec:3} hold also for $b=0$ which is a model with a cosmological constant. This means that one could add to the statement of the authors: ``data indicate that currently there is no compelling evidence for non-standard gravity'', that the data also indicate that currently there is no compelling evidence against non-standard gravity, since a model with $b=0$ but $d\neq 0$ would give the exact same results as the $\Lambda$CDM model. This highlights the point made earlier, that the presence of positive powers of $R$ in the action cannot be constrained using observational data relevant to the late time evolution. Such terms, however, could be seriously constrained using other types of data, such as the evolution of perturbations during inflation. 

\section{Conclusions}
\label{sec:5}

Starting by setting up the formalism needed, a model with both positive and negative powers of the scalar curvature, additional to the standard Einstein--Hilbert term in the gravitational action, was studied. The exponents and the coefficients of the extra terms were left arbitrary and all eras of cosmological evolution were studied. It was shown that, even though such a model can lead to early-time inflation at large values of the scalar curvature and late-time expansion at small values, as shown in \cite{sot2}, it is not possible to have a smooth passage from the former to the latter for any model. If the curvature dominates the expansion there seems to be no turning back to ordinary matter domination, so it is reasonable to require that this should only happen at late times. Avoiding such an inflationary regime however, does not necessarily imply that additional positive powers of the curvature should not be present, but merely that they should be seriously constrained. In that case one can also have the usual scenario where inflation is driven by an inflaton field. Other constraints were derived as well, for the coefficients of the positive power terms, by requesting that the modified Friedmann equation does not differ significantly from the standard one during big bang nucleosynthesis and that the correct Newtonian limit is obtained. Using the above constraints, it was shown that the late time evolution of the universe is not affected by the presence of additional positive power terms of the scalar curvature in the action.

As a conclusion, any result derived for a model with only negative powers of the curvature using late time observational data, as done in \cite{ama} is general enough to be applied to a model including additional positive power terms as well. At the same time, however, since such tests are totally insensitive to the presence of positive power terms, they cannot be used to constrain them and therefore favour or disfavour their presence. To close I would also like to mention the
following. The results of \cite{ama} seem to disfavour the presence of a negative power of the curvature in the gravitational action by showing that $\Lambda$CDM is the best fit model. However, the model that better fits the data, does not necessarily have to be the physically preferred one, unless it can also be theoretically motivated and justified. The $\Lambda$CDM model comes with a burden, known as the coincidence problem, and this is the motivation for creating the numerous alternative models present in the literature.  In this sense, what we are interested in is not whether a model can fit the data better than the $\Lambda$CDM one, but if we can produce a model which is theoretically motivated or at least theoretically explained and at the same time fits the current data reasonably well. Whether this is true for $f(R)$ theories of gravity in the Palatini formalism is a question that remains unanswered.

\section*{Acknowledgments}
The author is grateful to Stefano Liberati for numerous enlightening discussions and valuable suggestions during the preparation of this work. He would also like to thank
John Miller for his critical reading of the manuscript and for constructive 
comments and Urbano Fran\c{c}a for stimulating conversations on the subject of this paper.

\end{document}